\documentclass[a4paper,twoside,11pt]{article}

\usepackage{cmap}               
\usepackage[T2A]{fontenc}       
\usepackage[cp1251]{inputenc}   
\usepackage[russian]{babel}     
\usepackage{fancyhdr}
\usepackage{newprog1e}

\usepackage{indentfirst}        

\usepackage[unicode]{hyperref}

\usepackage{ccaption}
\captiondelim{ }

\usepackage{amssymb} 
\renewcommand*{\leq}{\leqslant}

\renewcommand*{\geq}{\geqslant}

\usepackage{amsmath}
\usepackage{float}
\usepackage{algorithm2e}
\usepackage{tikz}
\usetikzlibrary{shapes,arrows,matrix,patterns}

\rusabstr{В статье рассматривается метод реализации табличного элемента управления (грида) для отображения записей из базы данных, способного быстро отображать записи, соответствующие позиции бегунка полосы прокрутки, и быстро переходить к записи, заданной по комбинации ключевых полей.}
\rubrika{}
\curyear{2016}
\journalnumber{??}
\dateinput{??}
\author{\bfseries{Пономарев И.~Н.*}\\ 
\itshape{*Московский физико-технический институт}\\
\itshape{E-mail: \href{mailto:iponomarev@mail.ru}{iponomarev@mail.ru}}} 
\title{Реализация быстродействующего табличного элемента управления для отображения записей базы данных}
\date{}

\authorlist{ПОНОМАРЕВ И.~Н.}
\titlehead{РЕАЛИЗАЦИЯ ТАБЛИЧНОГО ЭЛЕМЕНТА УПРАВЛЕНИЯ}
\headerdef

\udk{004.42}

\begin{document}

\maketitle


\subsection*{Введение. Постановка задачи}

Классическая работа с полосой прокрутки таблицы базы данных предполагает поддержку двух ключевых операций: отображение записей, соответствующих положению бегунка полосы прокрутки, и переход к записи, заданной по комбинации ключевых полей.

Из-за большого числа записей их полная загрузка в оперативную память бывает недопустима, а перекладывание указанных операций на СУБД ведёт к серьёзным проблемам производительности. Операции типа select count(*) и select ... offset ... являются медленными, т. к. предполагают перебор записей таблицы. В результате часто при работе с таблицами происходит либо отказ от использования таблицы (использование постраничного отображения), либо пользователю тем или иным способом предоставляется лишь иллюзия того, что с помощью бегунка полосы прокрутки можно быстро перейти к любой из записей.

Однако в некоторых случаях пользователю требуется именно классическая работа с бегунком, т. е. возможность прямо перейти с помощью полосы прокрутки к любой части таблицы. Алгоритм, обеспечивающий такую работу грида при сохранении быстродействия, мы рассматриваем в настоящей работе.

Ограничивающими условиями для нас будут следующие: 1) набор данных, отображаемых в таком гриде, может быть отсортирован не произвольным образом, а только лишь по индексированному набору полей, 2) если в составе индекса содержится строковое поле, алгоритм должен быть <<обучен>> правилам сопоставления базы данных (collation rules) для всех символов, которые могут встретиться в данном строковом поле.

\subsection*{<<Быстрые>> и <<медленные>> запросы к СУБД}

Как уже говорилось, быстрыми запросами к СУБД являются запросы, не использующие переборы записей, т. е. такие, в которых получение записей производится поиском по индексу. При условии, что по полю \texttt{key} построен индекс, быстрыми являются следующие запросы:

\begin{enumerate}
	\item Найти первую и последнюю запись в наборе данныx (запрос A):
			
			\texttt{select ... order by key [desc] limit 1}

	\item Найти $h$ первых записей с ключом, большим или равным данному значению $K$ (запрос B): 

      \texttt{select ... order by key where key >= K limit h}

\end{enumerate}

Эти запросы можно обобщить на случай сортировки по набору из нескольких полей \texttt{order by key1, key2, key3...} при условии, что на этих полях построен составной индекс. Условие $k \geq K$ должно быть обобщено на логическое условие сравнения нескольких значений в лексикографическом порядке:

\begin{equation}
\label{eq:lex1}
\begin{split}
k_1 > K_1 \vee (k_1 = K_1 \wedge k_2 \geq K_2) \\
\vee (k_1 = K_1 \wedge k_2 = K_2 \wedge k_3 \geq K_3) \vee \ldots
\end{split}
\end{equation}

На практике тонкость заключается в том, что запрос с выражением вида \texttt{where k1 > K1 or (k1 = K1 and k2 >= K2)} может не выполняться за быстрое время: например, выполнение такого запроса в СУБД PostgreSQL с анализом плана выполнения показывает, что система не в состоянии использовать имеющийся индекс и использует вместо этого сканирование таблицы (table scan). Причиной тому выражение \texttt{or} на верхнем уровне логического условия, которое плохо поддаётся оптимизации. Проблема решается заменой условия \texttt{where} на логически эквивалентную формуле \eqref{eq:lex1} формулу 

\begin{equation}
\label{eq:lex2}
\begin{split}
k_1 \geq K_1 \wedge (k_1 > K_1 \vee \\(k_2 \geq K_2  \wedge (k_2 > K_2 \vee \ldots)))
\end{split}
\end{equation}

При таком ограничивающем условии СУБД распознаёт возможность применения составного индекса.

Не являются быстрыми следующие запросы:
\begin{enumerate}
	\item Подсчитать общее количество записей в наборе данных (запрос C):
	
	   \texttt{select count(*) ...}
    
	\item Подсчитать число записей, предшествующих записи с ключом, большим или равным данному (запрос D):
	
		    \texttt{select count(*) ... where key < K}
\end{enumerate}

В общем случае системе понадобится перечислить все попадающие в фильтр записи для их подсчёта, поэтому на эти запросы затрачивается время, пропорциональное общему количеству записей.

Основная идея нашего подхода заключается в том, чтобы в процедурах, требующих быстрого отклика для пользователя, использовать только быстрые запросы к СУБД. 


\subsection*{Зависимость ключа и номера записи} 

Для начала допустим, что интересующий нас набор данных отсортирован по одному единственному целочисленному полю (далее это ограничение будет снято). 

Сопоставим каждому положению бегунка полосы прокрутки целочисленное значение $\lambda$, обозначающее количество записей <<сверху>> от границы отображаемого окна, или, иначе говоря, число <<пропущенных перед началом вывода>> записей. Таким образом, при $\lambda = 0$ отображаются записи с самого начала, при $\lambda = 1$ записи отображаются, начиная со второй и т. д. Если в окно умещается $h$ записей, а всего в наборе $N$ записей, то значения полосы прокрутки должны изменяться в пределах от $0$ до $N-h$ включительно. 

Теперь рассмотрим функцию $f(k) = \lambda$, сопоставляющую каждому ключу $k$ результат выполнения запроса D, т. е. число записей, предшествующих записи с ключом $k$. Обратная ей функция $f^{-1}(\lambda)$ обладает тем свойством, что подстановка её результата в качестве параметра в быстро выполняющийся запрос B возвратит набор строк, в котором записи будут начинаться с $\lambda + 1$-й по счёту. Таким образом, если есть возможность за малое время вычислять функцию $f$, то задача перехода к записи, имеющей ключ $k$, сводится к вызову запроса B, а также вычислению $\lambda = f(k)$ и выставлению полосы прокрутки в положение $\lambda$. Если есть возможность за малое время вычислять $f^{-1}$, то отображение записей, соответствующих заданному положению полосы прокрутки, сводится к вычислению $k = f(\lambda)$ и вызову запроса B с параметром $k$.

Функция $f$ целиком определяется данными в таблице, поэтому, чтобы точно восстановить взаимозависимость значений ключа и номера записи, необходимо и достаточно прочесть из базы все записи \emph{через одну}: для корректного срабатывания запроса B в промежуточных точках можно считать, что $f^{-1}(2n+1) = f^{-1}(2n)+1$, $n = 0, 1\ldots$). На практике без чтения большого количества данных можно обойтись: как мы покажем, значения $f$ могут быть достаточно точно приближены при помощи кусочной интерполяции по относительно малому числу промежуточных точек.

Допустим, что при помощи запросов A и D нам стали известны минимальное и максимальное значения ключа $k_{\min}$ и $k_{\max}$, а также значение $\lambda_{\max}$ (равное, естественно, числу записей в таблице минус единица). Тот факт, что запрос D --- <<медленный>>, не играет существенной роли, что будет видно из дальнейшего. Помимо значений в крайних точках $f(k_{\min}) = 0$ и $f(k_{\max}) = \lambda_{\max}$, про функцию $f$ нам известны следующие факты:

\begin{enumerate}
	\item $f(k)$ монотонно растёт, 
	\item при увеличении $k$ на единицу, $\lambda$ увеличивается на единицу или не увеличивается, поэтому график функции $f$, кроме точки $(k_{\min}, 0)$, целиком лежит в параллелограмме $(k_{\min} + 1, 1)$, $(k_{\min}+\lambda_{\max}, \lambda_{\max})$,  $(k_{\max}, \lambda_{\max})$, $(k_{\max} - \lambda_{\max} + 1, 1)$,
	\item общее число возможных функций $f$ равно числу способов распределения $\lambda-1$ записей по $k_{\max} - k_{\min} - 1$ значениям ключа (позиции первой и последней записи фиксированы), т.~е. 
	\begin{equation}\label{eq:totalfuncnum}F = \binom{k_{\max} - k_{\min} - 1}{\lambda_{\max} - 1}\end{equation}
	\item число возможных функций $f$, которые в точке $k$ принимают значение $\lambda$, определяется произведением количества комбинаций записей с ключом, меньшим $k$, и большим или равным $k$:
\end{enumerate}

\begin{equation}\label{eq:funcnum}F_{k,\lambda}=\binom{k - k_{\min} - 1}{\lambda - 1}\binom{k_{\max} - (k - k_{\min})}{ \lambda_{\max} - \lambda}\end{equation}

Если каждый из возможных вариантов считать равновероятным, то вероятность того, что для заданного значения $k$ имеется ровно $\lambda$ записей с ключом, строго меньшим $k$, задаётся отношением $F_{k,\lambda}/F$, являющимся гипергеометрическим распределением вероятностей.

В случае $k=k_{\min}$ всегда $\lambda = 0$. На отрезке $k=(k_{\min} + 1)\ldots k_{\max}$ среднее значение $\lambda$ определяется формулой (см. напр. \cite{F2011}):
\begin{equation}
\label{eq:meanlambda}
\overline{\lambda} = \frac{(\lambda_{\max} - 1)(k - k_{\min} - 1)}{k_{\max} - k_{\min} - 1} + 1
\end{equation}

Дисперсия значения  $\lambda$, по \cite{F2011}, имеет форму перевернутой параболы с нулями на краях отрезка $(k_{\min} + 1)\ldots k_{\max}$ и максимумом посередине:
\begin{equation}
\label{eq:variance}
\begin{split}
D_\lambda = \frac{(\lambda_{\max} - 1)(k - k_{\min} - 1)(k_{\max} - k)}{(k_{\max} - k_{\min} - 1)^2}\\
\times \frac{(k_{\max} - k_{\min} - \lambda_{\max})}{(k_{\max} - k_{\min} - 2)}
\end{split}
\end{equation}

На рис.~\ref{fig:combinations} показаны границы возможных значений функции, её среднее значение для всех комбинаций, а также границы среднеквадратичного отклонения при $k_{\max} - k_{\min} = 60$, $\lambda_{\max} = 6$.

\begin{figure}[ht]
\centering 
\begin{tikzpicture}[x=0.1cm,y=0.8cm]
\draw [thick, <->] (0,7) -- (0,0) -- (70,0);
\node [below right] at (70,0) {$k$};
\node [left] at (0,7) {$\lambda$};
\draw[thick] (0, 0) -- (6, 6) -- (60, 6) -- (55, 1) -- (1, 1);
\draw[thick] (1, 1) -- (60, 6);
\draw [pattern=north west lines] (1,1) -- (10,1) --  (11,1.04) -- (13,1.15) -- (15,1.27) -- (19,1.53) -- (21,1.67) -- (23,1.82) -- (27,2.13) node[below right] {$\overline{\lambda} - \sqrt{D_\lambda}$} -- (29,2.3) -- (31,2.46) -- (35,2.82) -- (37,3) -- (39,3.19) -- (43,3.58) -- (45,3.79) -- (47,4) -- (49,4.23) -- (50,4.34) -- (51,4.46) -- (53,4.71) -- (55,4.98) -- (57,5.27) -- (59,5.64) -- (60,6) -- (1, 1)
 -- (2,1.36) -- (4,1.73) -- (6,2.02) -- (8,2.29) -- (10,2.54) -- (11,2.66) -- (13,2.89) -- (15,3.1) -- (19,3.52) -- (21,3.72) -- (23,3.91) -- (27,4.27) -- (29,4.45) -- (31,4.62) node[above left] {$\overline{\lambda} + \sqrt{D_\lambda}$}  -- (35,4.95) -- (37,5.1) -- (39,5.25) -- (43,5.54) -- (45,5.67) -- (47,5.79) -- (49,5.91) -- (50,5.96) -- (51,6) -- (60,6);
\draw[fill] (0,0) circle [radius=1pt];
\draw[fill] (1,1) circle [radius=1pt];
\draw[fill] (60,6) circle [radius=1pt];

\draw[fill] (0,1) circle [radius=1pt];
\draw[fill] (0,6) circle [radius=1pt];
\draw[fill] (1,0) circle [radius=1pt];
\draw[fill] (55,0) circle [radius=1pt];
\draw[fill] (60,0) circle [radius=1pt];

\draw[->] (20, 4)node[left] {$\overline{\lambda}$} to[out=0,in=140] (36.5, 4);
\draw[dashed] (60, 6) -- (60,0)node[below right]{$k_{\max}$};
\draw[dashed] (6, 6) -- (0,6)node[left]{$\lambda_{\max}$};
\draw[dashed] (1, 1) -- (0,1)node[left]{$1$};
\draw[dashed] (1, 1) -- (1,0)node[below right]{$k_{\min} + 1$};
\draw[dashed] (55, 1) -- (55,0)node[below left]{$k_{\max} - \lambda_{\max} + 1$};
\end{tikzpicture}
\caption{Зависимость ключа и номера записи: допустимые границы, среднее значение, среднеквадратичное отклонение\label{fig:combinations}}
\end{figure}

Обратная к \eqref{eq:meanlambda} функция, в соответствии с \cite{F2011}, даёт несмещенную оценку с минимальной дисперсией для значения $k$:

\begin{equation}\label{eq:meank}
k = \frac{(\lambda - 1)(k_{\max} - k_{\min} - 1)}{\lambda_{\max} - 1} +k_{\min} + 1
\end{equation}

Формулы \eqref{eq:meanlambda} и \eqref{eq:meank} мы примем в качестве оценки $f$ и $f^{-1}$ для $k=(k_{\min} + 1)\ldots k_{\max}$, а в точке $k_{\min}$ заведомо $f = 0$. 

Если в дальнейшем для какого-то $k'$, $k_{\min} < k' < k_{\max}$ мы узнаём новое точное значение $0 < \lambda' < \lambda_{\max}$, мы можем добавить пару $(k', \lambda')$ к интерполяционной таблице и получить уточнённый расчёт значения $k(\lambda)$ для решения задачи прокрутки, а применяя обратную интерполяцию, вычислять уточнённое значение $\lambda(k)$ при решении задачи позиционирования. Разумеется, все эти операции можно реализовать так, чтобы они работали за время, логарифмическое по количеству точек в интерполяционной таблице.

Использование кусочно-линейной интерполяции для поиска записей в таблице лежит в основе алгоритма интерполяционного поиска, исследованного, например, в \cite{P1978}. В частности, там приводится оценка сверху для средней ошибки значения $\lambda$ (менее чем $\frac{1}{2}\sqrt{\lambda_{\max}}$), которую легко вывести из \eqref{eq:variance}.

\subsection*{Обобщение на практически встречающиеся случаи} На практике дело не ограничивается единственным целочисленным полем для сортировки набора данных. Во-первых, тип данных может быть другим (строка, дата,\ldots) Во-вторых, сортируемых полей может быть несколько. Это затруднение устраняется, если мы умеем вычислять 

\begin{enumerate}
	\item функцию-нумератор $g(K_1,\ldots K_n)=\kappa$, переводящую набор значений полей произвольных типов в натуральное число, 
  \item обратную ей функцию $g^{-1}(\kappa)=(K_1,\ldots K_n)$, переводящую натуральное число обратно в набор значений полей, $g^{-1}g(K_1,\ldots K_n)=(K_1,\ldots K_n)$.
\end{enumerate}

Функция-нумератор должна обладать тем свойством, что если набор $(K_1,\ldots K_n)$ меньше набора $(K'_1,\ldots K'_n)$ в лексикографическом смысле (см. формулы~\eqref{eq:lex1} и~\eqref{eq:lex2}), то должно быть \begin{equation}g(K_1,\ldots K_n) < g(K'_1,\ldots K'_n).\end{equation} 

Для представления значений $\kappa$ не подходят стандартные 32- и 64-битовые целочисленные типы: так, чтобы перенумеровать одни лишь всевозможные 10-байтовые строки, уже не хватит 64-битового (8-байтового) целого. В своей реализации мы использовали класс java.math.BigInteger из стандартной библиотеки Java, способный представлять целые числа произвольной величины. При этом объём оперативной памяти, занимаемой значением $\kappa$, примерно равен объёму, занимаемому набором значений $K_1,\ldots K_n$.

Говоря языком математики, биекция $g$ должна устанавливать изоморфизм порядка между множеством возможных значений наборов полей и множеством натуральных чисел. 

При наличии обратимой функции-нумератора $g$ и обратимой функции-интерполятора $f$, 

\begin{itemize}
\item \textbf{прокрутка} грида сводится к вычислению значений ключевых полей $(K_1,\ldots K_n)=g^{-1}(f^{-1}(\lambda))$, где $\lambda$ -- положение вертикальной полосы прокрутки, после чего быстрый запрос к БД находит $h$ первых записей, больших или равных $(K_1,\ldots K_n)$,
\item \textbf{позиционирование} сводится к считыванию $h$ первых записей из БД по заранее известным значениям $(K_1,\ldots K_n)$, и к вычислению положения бегунка полосы прокрутки $\lambda = f(g(K_1,\ldots K_n))$.
\end{itemize}
\subsection*{Общая схема взаимодействия процедур}
Общая схема взаимодействия процедур системы показана на рис.~\ref{fig:scheme}. Cплошной стрелкой показана последовательность выполнения процедур, пунктирной стрелкой~--- асинхронный вызов в отдельном потоке выполнения.
 
\begin{figure*}[t]
\centering
\begin{tikzpicture}[
     block/.style={rectangle, draw=black, thick, fill=white, align=left, text ragged, minimum width=8em,rounded corners=2mm},
		 line/.style={draw, thick, -latex', shorten >=0pt},
		 bullit/.style={circle, draw=black, thick, fill=white, radius=5mm},
		 blackbullit/.style={circle, draw=black, thick, fill=black, radius=5mm}]
	  \matrix[column sep=3mm,row sep=5mm] {
		& \node (h2) {Прокрутка}; & \node  (h3) {Позиционирование}; &  \\      
    & \node [blackbullit] (s0) {}; & \node [bullit] (p0) {}; &  \\                           
      \node (h5)[align=left] {Запрос к БД};   
			& \node [block] (s1) {Вернуть \\ ближайшие к \\ $(K_1,\ldots K_n)$ \\ $h$ записей, \\ $(K_1,\ldots K_n) \leftarrow$ \\ ключи  первой \\ из них }; 
			& \node [block] (p1) {Вернуть \\ $h$ записей, \\ начиная с \\ $(K_1,\ldots K_n)$ };
			&  \node [block] (r1)  {$\lambda \leftarrow$ число \\ записей, меньших \\  $(K_1,\ldots K_n)$ \\ \textbf{(медленный} \\ \textbf{запрос)} };  \\
      \node (h6)[align=left] {Нумератор};     
			& \node [block] (s2) {$(K_1,\ldots K_n) \leftarrow g^{-1}(\kappa)$}; 
			& \node [block] (p2) {$\kappa \leftarrow g(K_1,\ldots K_n)$ }; 
			&  \node [block] (r2)  {$\kappa \leftarrow g(K_1,\ldots K_n)$ };  \\
      \node (h7)[align=left] {Интерполятор};  
			& \node [block] (s3) {$\kappa \leftarrow f^{-1}(\lambda)$};       
			& \node [block] (p3) {$\lambda \leftarrow f(\kappa)$};         
			&  \node [block] (r3)  {вставить новую \\ интерполяционную \\ точку $(\kappa, \lambda)$};  \\    
			\node (h8)[align=left] {Полоса \\ прокрутки}; 
			& \node [block] (s4) {$\lambda\leftarrow$ позиция \\ полосы прокрутки};       
			& \node [block] (p4) {позиция полосы \\ прокрутки $\leftarrow\lambda$}; 
			&  \node [block] (r4)  {позиция полосы \\прокрутки $\leftarrow\lambda$, \\<<отскок>>};  \\   
      & \node [bullit] (s5) {}; 
			& \node [blackbullit] (p5) {};  
			&  \node [blackbullit] (r5) {}; \\ 
		};
      \path[line] (s5) -- (s4);
      \path[line] (s4) -- (s3);
      \path[line] (s3) -- (s2);
      \path[line] (s2) -- (s1);
      \path[line] (s1) -- (s0);
      \path[line] (p0) -- (p1);
      \path[line] (p1) -- (p2);
      \path[line] (p2) -- (p3);
      \path[line] (p3) -- (p4);
      \path[line] (p4) -- (p5);
			\path[line] (r1) -- (r2);
      \path[line] (r2) -- (r3);
      \path[line] (r3) -- (r4);
			\path[line] (r4) -- (r5);
			
			\path[line] (s1)[dashed] edge [bend left = 30] (r1);
			\path[line] (p1)[dashed] -- (r1);
\end{tikzpicture}
\caption{Общая схема взаимодействия процедур.\label{fig:scheme}}
\end{figure*}

Допустим, что пользователь изменил положение бегунка вертикальной полосы прокрутки (см. левый нижний угол диаграммы рис.~\ref{fig:scheme}). 

Интерполятор вычисляет значение номера комбинации значений ключевых полей ($\kappa=f^{-1}(\lambda)$) с типом BigInteger. На основе этого значения нумератор восстанавливает комбинацию ключевых полей $(K_1,\ldots K_n)=g^{-1}(\kappa)$. Важно понимать, что на данном этапе в полях $K_1,\ldots K_n$ не обязательно будут находиться значения, действительно присутствующие в базе данных: там будут лишь приближения. В строковых полях, скорее всего, будет бессмысленный набор символов. Тем не менее, вывод из базы данных $h$ строк с ключами, большими или равными набору $K_1,\ldots K_n$, окажется приблизительно верным результатом для данного положения полосы прокрутки.

Если пользователь отпустил полосу прокрутки, асинхронно (в отдельном потоке выполнения) запускается запрос к БД, определяющий порядковый номер записи, а значит, и точное положение полосы прокрутки, которое соответствует тому, что отображено пользователю. Когда запрос будет завершён, на основе полученных данных будет пополнена интерполяционная таблица. Кроме того, если на экране пользователя к тому моменту ничего не изменится, бегунок полосы прокрутки <<отскочит>> на новое, уточнённое положение.

При переходе к записи последовательность вызовов процедур происходит в обратную сторону. Т. к. значения ключевых полей уже известны, для пользователя сразу извлекаются данные из базы. Нумератор вычисляет $\kappa  = g(K_1,\ldots K_n)$, и затем интерполятор определяет приблизительное положение полосы прокрутки как $\lambda = f(\kappa)$. Параллельно, в асинхронном потоке выполнения, выполняется уточняющий запрос, по результатам которого в интерполяционную таблицу добавляется новая точка. Если на экране пользователя к тому моменту ничего не изменится, бегунок полосы прокрутки <<отскочит>> на новое, уточнённое положение.
 
\subsection*{Реализация интерполятора}

Объект-интерполятор должен хранить в себе промежуточные точки монотонно растущей функции между множеством 32-битных целых чисел (номеров записей в таблице) и множеством объектов типа BigInteger (порядковых номеров комбинаций значений ключевых полей). 

Сразу же после инициализации грида необходимо в отдельном потоке выполнения запросить общее количество записей в таблице, чтобы получить корректное значение $\lambda_{\max}$. До того момента, как это значение будет получено при помощи выполняющегося в параллельном потоке запроса, можно использовать некоторое значение по умолчанию (например, 1000)~-- это не повлияет на корректность работы.

Интерполятор должен уметь за быстрое по количеству интерполяционных точек время вычислять значение как в одну, так и в другую сторону. Однако заметим, что чаще требуется вычислять значение порядкового номера комбинации по номеру записи: такие вычисления производятся много раз за секунду в процессе прокрутки грида пользователем. Поэтому за основу реализации модуля интерполятора удобно взять словарь на основе бинарного дерева, ключами которого являются номера записей, а значениями -- порядковые номера комбинаций (класс TreeMap<Integer, BigInteger> в языке Java).

Ясно, что по заданному номеру $\lambda$ такой словарь за логарифмическое время находит две точки ($\underline{\lambda} \leq \lambda \leq \overline{\lambda}$), между которыми строит интерполяцию по формуле~\eqref{eq:meank}. Но тот факт, что функция растёт монотонно, позволяет за быстрое время производить и обратное вычисление. В самом деле: если дан номер комбинации $\kappa$, $\kappa_{\min} \leq \kappa \leq \kappa_{\max}$, поиск кусочного сегмента, в котором лежит $\kappa$, можно произвести в словаре методом дихотомии. Отыскав нужный сегмент, мы производим обратную интерполяцию по формуле~\eqref{eq:meanlambda} и находим номер $\lambda$, соответствующий $\kappa$.

При пополнении словаря интерполяционными точками необходимо следить за тем, чтобы интерполируемая функция оставалась монотонной. Так как другие пользователи могут удалять и добавлять записи в просматриваемую таблицу, актуальность известных словарю интерполяционных точек может утратиться, а вновь добавляемая точка может нарушить монотонность. Поэтому метод добавления новой интерполяционной точки должен проверять, что <<точке слева>> от только что добавленной соответствует меньшее, а <<точке справа>> -- большее значение. Если оказывается, что это не так, следует исходить из предположения, что последняя добавленная точка соответствует актуальному положению вещей, а некоторые из старых точек утратили свою актуальность. По отношению к вновь добавленной точке следует удалять все точки слева, содержащие большее значение, и все точки справа, содержащие меньшее значение (см. рис.~\ref{fig:interp}).

\begin{figure}[ht]
\centering
\begin{tikzpicture}[x=0.4cm,y=0.4cm]
\draw [thick, <->] (0,8) -- (0,0) -- (8,0);
\node [below right] at (8,0) {$\lambda$};
\node [left] at (0,8) {$\kappa$};
\draw[fill] (1,1) circle [radius=1pt];
\draw[fill] (3,4) circle [radius=1pt];
\draw[fill] (5,6) circle [radius=1pt];
\draw[fill] (7,7) circle [radius=1pt];
\draw[fill] (7,0) circle [radius=1pt];
\draw[thick] (0,0) -- (1,1) -- (3,4) -- (5,6) -- (7,7);
\draw[dashed] (7, 7) -- (7,0);
\node [below] at (7,0) {$\lambda_{\max}$};
\draw [thick, <->] (10,8) -- (10,0) -- (18,0);
\node [below right] at (18,0) {$\lambda$};
\node [left] at (10,8) {$\kappa$};
\draw[fill] (11,1) circle [radius=1pt];
\draw[fill] (14,3) circle [radius=1pt];
\draw[fill] (15,6) circle [radius=1pt];
\draw[fill] (17,7) circle [radius=1pt];
\draw[thick] (10,0) -- (11,1) -- (14,3) -- (15,6) -- (17,7);
\draw[thick,dashed] (11, 1) -- (13,4) -- (14,3);
\draw[thick,fill=white] (13,4) circle [radius=1.5pt];
\draw[dashed] (17, 7) -- (17,0);
\draw[fill] (17,0) circle [radius=1pt];
\node [below] at (17,0) {$\lambda_{\max}$};
\end{tikzpicture}
\caption{Удаление точки, нарушающей монотонность, при вставке новых данных в интерполяционную таблицу.\label{fig:interp}}
\end{figure}

Также интерполятор должен содержать в себе механизм, в целях экономии памяти защищающий словарь от переполнения излишними точками, и отбрасывающий наименее существенные из них.


\subsection*{Нумераторы для числовых типов данных}

Назовём мощностью машинного типа данных количество различных значений, которые представимы при помощи этого типа.

Мощность типа BIT равна 2, нумерация его значений тривиальна: $\mathrm{false} \leftrightarrow 0$, $\mathrm{true} \leftrightarrow 1$.

Мощность типа INT (32-битовое целое со знаком) равна $2^{32}$. INT-значение есть число между $-2147483648$ и $2147483647$. Таким образом, нумератор для типа INT есть просто $g(k) = k + 2147483648$ (конечно, выполнять сложение следует, уже приведя $k$ к типу BigInteger). Обратная функция $g^{-1}(\kappa) = \kappa - 2147483648$.

Подобным же образом можно построить нумератор и для 64-битовых целых чисел со знаком.

Числа с типом DOUBLE (двойной точности с плавающей точкой), представленные в формате IEEE 754, обладают тем свойством, что их можно (за несущественными исключениями вроде NaN и $\pm0$) сравнивать как целые 64-битовые числа со знаком. В языке Java получить для значения с типом double его 64-битовый образ в формате IEEE 754 можно с помощью метода Double.doubleToLongBits.

Наконец, значения DATETIME, определяющие момент времени с точностью до миллисекунды, также могут быть сведены к 64-битовому целому числу со знаком, задающему так называемое <<UNIX-время>>, т. е. количество миллисекунд от полуночи 1 января 1970 года. В языке Java это делается при помощи метода Date.getTime.

Методы реализации нумераторов для строковых (VARCHAR(m)) типов и составных ключей рассмотрены далее.

\subsection*{Нумератор для составных ключей}

Пусть типы данных составного ключа имеют мощности $N_1,\ldots, N_n$. Тогда общее количество возможных комбинаций значений ключевых полей равно $N_1N_2\ldots N_n$. Если вычислены функции нумераторов для значения каждого из полей, $\kappa_i$ -- порядковый номер значения $i$-го поля, то функция нумератора составного ключа может быть представлена как

\begin{equation}\label{eq:comp1}
\begin{split}
g(K_1,\ldots K_n) =  \kappa_n + N_n\kappa_{n-1} \\
+ N_nN_{n-1}\kappa_{n-2} + \ldots.
\end{split}
\end{equation}

Значение $\kappa_1$  имеет наибольший вес, $\kappa_n$ -- наименьший. Также легко проверить, что $g(N_1 - 1, N_2-1,\ldots N_{n-1}-1) = N_1N_2\ldots N_n - 1$.

Вычисление $g$ напрямую по формуле \eqref{eq:comp1} требует $(n-1)n/2$ операций умножения. Сократить количество умножений до $n-1$ при том же количестве сложений можно, воспользовавшись аналогом схемы Горнера:

\begin{equation}\label{eq:comp2}
\begin{split}
g(K_1,\ldots K_n) = ((\ldots(\kappa_1N_2 + \kappa_2)\ldots)N_{n-1} \\
+ \kappa_{n-1})N_n + \kappa_n.
\end{split}
\end{equation}

Вычислить обратную функцию $g^{-1}$, получив из значения $g$ массив значений $\kappa_i$, можно по следующему простому алгоритму:

\begin{algorithm}
	$i \leftarrow n$\;
	\While {$i>0$} {
	  $\kappa_i \leftarrow g \mod N_i$\; 
	  $g \leftarrow \left\lfloor g / N_i \right\rfloor$\;
		$i \leftarrow i - 1$\;
	}
\end{algorithm}

\subsection*{Нумератор для строк (простой лексикографический порядок)}

Сперва заметим, что если дан алфавит из $a$ символов, то общее количество строк длины не более $m$ в этом алфавите равно

\begin{equation}\label{eq:str1}
1 + a + a^2 + \ldots + a^{m} = \frac{a^{m+1}-1}{a-1}.
\end{equation}

Здесь единица соответствует пустой строке, $a$ -- количество строк из одного символа, $a^2$~-- количество строк из двух символов и т.~д., а в итоге получается сумма геометрической прогрессии.

Представим теперь произвольную строку $c$ длины $l \leq m$ как массив $(c_0, c_1,\ldots c_{l-1})$, где $c_i$ -- номер $i$-го символа строки в алфавите (считая с нуля), позиции символов в строке тоже считаем с нуля. Тогда строка $c$ в простом лексикографическом порядке будет иметь номер 
\begin{equation}\label{eq:str2}
\begin{split}
g(c) = l + \frac{a^m - 1}{a - 1}c_0 + \frac{a^{m -1} - 1}{a - 1}c_1 + \ldots \\
+ \frac{a^{m - l + 1} - 1}{a - 1}c_{l-1}.
\end{split}
\end{equation}

Докажем формулу \eqref{eq:str2} индукцией по $m$ и $l$. Для иллюстрации примем, что алфавит состоит всего из двух букв: a и b.

Если $m = 0$, то единственный вариант -- это пустая строка с номером 0.

Если $m = 1$, то пустая строка будет иметь номер 0, а каждая односимвольная будет иметь номер $1 + c_0$. Единица прибавляется потому, что при сравнении строк меньше любой односимвольной строки будет пустая строка: для двухсимвольного алфавита $0 \leftrightarrow \textnormal{`'}$, $1 \leftrightarrow \textnormal{`a'}$, $2 \leftrightarrow \textnormal{`b'}$.

Если $l \leq 1$, $m \geq 1$, то по-прежнему $0 \leftrightarrow \textnormal{`'}$, $1 \leftrightarrow \textnormal{`a'}$. Но теперь между строками `a' и `b' в пространстве лексикографически отсортированных строк находятся все возможные строки вида <<`a' плюс любая строка длиной не более $m - 1$>>: a, aa, aaa\ldots, aab\ldots:

\[
\overbrace{a\underbrace{\square\square\square \ldots \square}_{\leq m-1}}^{\leq m}.
\]

Количество таких строк равно, по \eqref{eq:str1}, $(a^m-1)/(a-1)$, и окончательно для односимвольных строк

\[
g(c) = 1 + \frac{a^m - 1}{a - 1}c_0.
\]

Пусть к строке добавляется ещё один символ, при этом по-прежнему $l\leq m$. В качестве последнего слагаемого к \eqref{eq:str2} добавляется номер этого символа с соответствующим весом (равным числу строк длиной $m - l$). К первому слагаемому в \eqref{eq:str2} для каждого дополнительного символа добавляется единица, за счёт того, что строка, полученная отбрасыванием последнего символа, будет в лексикографическом порядке меньше любой из строк длиной $l$. Формула \eqref{eq:str2} доказана.

Для оптимизации вычисления $g$ по формуле \eqref{eq:str2} и для вычисления обратной функции необходимо заранее заготовить массив коэффициентов 

\begin{equation}\label{eq:qi}
q_i = \frac{a^{m - i} - 1}{a - 1}, i = 0,\ldots m-1.
\end{equation}

Разумеется, пользоваться формулой \eqref{eq:qi} напрямую при заготовке значений $q_i$ не нужно: сэкономить на арифметических операциях можно, заметив, что все $q_i$ являются частичными суммами геометрической прогрессии, которую можно вычислять <<на ходу>> при заполнении массива $q_i$.

Алгоритм для вычисления обратной функции $g^{-1}$, как и в случае нумератора составного ключа, является вариацией алгоритма преобразования числа в систему счисления с произвольным основанием. Необходимо только на каждом шаге перед получением очередного символа вычитать единицу, помня о первом слагаемом в~формуле~\eqref{eq:str2}:

\begin{algorithm}
$i \leftarrow 0$\;
\While {$i < m \wedge g > 0$} {
				$g \leftarrow g - 1$\;
				$c_i \leftarrow \left\lfloor g / q_i \right\rfloor $\;
				$g \leftarrow  g \mod q_i$\;
				$i \leftarrow i+1$;
			}
\end{algorithm}

\subsection*{Нумератор для строк с учётом правил сопоставлений}

Порядок, в котором база данных сортирует строковые значения, в действительности отличается от простого лексикографического и использует так называемые правила сопоставления (collation rules, \cite{U2015}).

Базой данных при сравнении строк с учётом этих правил каждый символ рассматривается в трёх аспектах: собственно символ, его регистр (case) и вариант написания (accent). Например, русская буква <<е>> в большинстве случаев рассматривается как имеющая два варианта написания, каждый из которых имеет два регистра: е, Е; ё, Ё. Это позволяет при сортировке, нечувствительной к варианту написания (accent insensitive), не различать <<е>> и <<ё>>, а при сортировке, нечувствительной к регистру (case insensitive), не различать строчные и заглавные буквы. При этом, что считать отдельной буквой, а что -- вариантом написания другой буквы, зависит от культурных традиций и может различаться даже в языках, использующих один и тот же алфавит. 


Общий алгоритм сравнения строк с учётом правил сопоставления следующий:

\begin{enumerate}
	\item Строки сравниваются посимвольно без учёта регистров и вариантов. Если обнаружено различие, возвращается результат (<<больше>> или <<меньше>>).
	\item Если сортировка accent sensitive, сравниваются номера вариантов каждого из символов. Если обнаружено различие, возвращается результат.
\item Если сортировка case sensitive, сравниваются регистры каждого из символов. Если обнаружено различие, возвращается результат.
\item Если выход из алгоритма не произошёл до сих пор -- строки равны.
\end{enumerate}


Таким образом, нам, во-первых, необходимо модифицировать алгоритм работы нумератора для строк таким образом, чтобы он учитывал правила сопоставления, а во-вторых, необходимо уметь задавать различные правила сопоставления, <<обучая>> грид работе с той или иной базой данных.

Первая из этих задач относительно проста с учётом уже полученных результатов. Всякое строковое значение необходимо рассматривать не как одномерный массив символов $c_i$, а как массив трёхкомпонентных значений $c_{ij}$, $0 \leq j \leq 2$. Здесь $c_{i0}$ -- номер $i$-го символа в алфавите, $c_{i1}$ -- его же вариант написания, $c_{i2}$ -- его же регистр. Тогда работу со строковым полем можно производить аналогично работе с составным ключом, состоящим из трёх полей.

Если известны $a_0$~-- количество символов в алфавите, $a_1$~-- максимальное число вариантов написания и $a_2$~-- максимальное число регистров (в известных нам языках $a_2 = 2$), то 

\begin{equation}\label{eq:coll1}
g = (k_0 a_1^m + k_1) a_2^m + k_2,
\end{equation}

где $k_0$ -- вычисленное по первым компонентам строки значение формулы \eqref{eq:str2} ($a = a_0$), 

\[
\begin{array}{l}
k_1 = a_1^{m-1} c_{01} + a_1^{m -2 } c_{11} + \ldots,\\
k_2 = a_2^{m-1} c_{02} + a_2^{m -2 } c_{12} + \ldots
\end{array}
\]

Обратную функцию легко построить, используя уже вышеизложенные принципы: сперва необходимо разделить $g$ на три компоненты $k_1$, $k_2$ и $k_3$, затем получить массив трёхкомпонентных значений, на основании которого восстанавливается исходная строка.

В стандартной библиотеке Java имеются абстрактный класс java.text.Collator и его реализация java.text.RuleBasedCollator. Назначением этих классов является сравнение строк с учётом разнообразных правил сопоставлений. Доступна обширная библиотека готовых правил. К сожалению, эти классы не пригодны для использования с какой-либо иной, чем сравнение строк, целью: вся информация о правилах сопоставлений инкапсулирована, и её невозможно получить, штатным образом используя системную библиотеку.

Поэтому для решения нашей задачи понадобилось создать интерпретатор правил сопоставлений самостоятельно. 

Эту задачу облегчило изучение класса RuleBasedCollator. Главной заимствованной идеей стал язык определения правил сортировки, формальное описание которого приведено в документации \cite{J2016}. Понять принцип работы этого языка проще всего, рассмотрев пример правила:

%
%
%
%
%
\texttt{<г,Г<д,Д<е,Е;ё,Ё<ж,Ж<з,З<и,И;й,Й<к,К<л,Л}

%

Следующие знаки являются служебными в языке правил сопоставления:
\begin{enumerate}
	\item < -- разделение символов,
	\item ; -- разделение вариантов написания,
	\item , -- разделение регистров.
\end{enumerate}

При помощи выражений, подобных вышеприведённому, можно определить правила, соответствующие различным сопоставлениям различных баз данных. Т. к. язык правил сопоставлений достаточно примитивен, для его разбора достаточно алгоритма, работающего как детерминированный конечный автомат. В итоге по заданному выражению правил мы получаем экземпляр класса, способный

\begin{enumerate}
  \item по заданным правилам получить значения $a = a_0$ (количество символов) для вычисления по формуле \eqref{eq:str2}, а также $a_1$ (максимальное число вариантов) и $a_2$ (максимальное число регистров) для вычисления по формуле \eqref{eq:coll1} и соответствующих обратных функций,
	\item по заданному символу определить три его компоненты (номер символа в алфавите, номер варианта, номер регистра),
  \item по заданной тройке компонентов определить символ.
\end{enumerate}

Это позволяет завершить реализацию нумератора для строковых значений.

\subsection*{Практическая реализация}
Грид по приведённым здесь принципам был реализован на языке Java с использованием PostgreSQL в качестве СУБД. В качестве нагрузочного тестового набора данных использовалась база данных КЛАДР \cite{K2016}, содержащая $1075429$ названий улиц населённых пунктов России, сортировка производилась по различным полям и их комбинациям. Тест продемонстрировал работоспособность изложенных здесь принципов. При постоянном перемещении бегунка вертикальной полосы прокрутки для пользователя создаётся полная иллюзия прокручивания всех записей в реальном времени, через малое время после окончания прокручивания (когда срабатывает уточняющий запрос и в интерполяционную таблицу добавляется ещё одна точка) позиция бегунка полосы прокрутки уточняется, перескакивая на небольшое расстояние. Позиционирование позволяет моментально отобразить нужные записи и сразу же приблизительно выставить бегунок полосы прокрутки, через малое время его позиция также уточняется. По мере работы с гридом и накопления интерполяционных точек, <<отскоки>> становятся всё менее и менее заметными.

\subsection*{Прокрутка на малый шаг}
Важным нюансом при практической реализации явилась необходимость отдельной обработки \emph{передвижения бегунка прокрутки на малый шаг}.

Прокрутка на одну строку вверх или вниз происходит при щелчке мышью на стрелки <<вверх>> и <<вниз>> вертикальной полосы прокрутки. При щелчке на свободное поле полосы прокрутки сверху или снизу от бегунка происходит прокрутка на фиксированное (малое) количество строк. В этих случаях пользователь ожидает сдвига всех видимых на экране строчек на фиксированное число позиций. Интерполятор, не набравший достаточно интерполяционных точек, может повести себя непредсказуемо, отбросив отображаемую пользователю картину слишком далеко назад или вперёд, и после уточнения позиция полосы прокрутки не будет соответствовать тому, что хотел пользователь.

В этом случае, однако, использование интерполятора и не оправдано. Если известен предыдущий набор значений ключевых полей, то получить одну предыдущую (или одну следующую) запись можно быстрым запросом к базе данных (см. (\ref{eq:lex1}) и (\ref{eq:lex2})). 
После извлечения этих данных, помимо отображения их пользователю, можно пополнить интерполяционную таблицу ещё одной точкой, \emph{не прибегая к~запросу на подсчёт записей}, т.~к. полученная запись имеет номер, отличающийся от предыдущей на известное значение.

\subsection*{Начальное заполнение интерполяционной таблицы}
Другим важным нюансом при практической реализации явилась необходимость заполнять интерполяционную таблицу данными до того, как пользователь начинает прокрутку грида.

Нет ничего удивительного в том, что номера комбинаций $\kappa$ на основе данных в реальной таблице распределяются на числовой прямой очень неравномерно. Поэтому, при недостаточном количестве точек в интерполяционной таблице, пользователь, сместив бегунок полосы прокрутки на некоторое расстояние, может получить после уточнения позиции сильный <<отскок>> вперёд или назад. В итоге реальная позиция просматриваемых данных переместится или намного дальше, или, наоборот, намного ближе, чем хотел пользователь.

Испытания показывают, что погрешность на 20-25\% от длины полосы прокрутки при позиционировании является психологически допустимой, но не более того. Поэтому после отображения грида пользователю желательно обеспечить, чтобы максимальная длина <<отскока>> составляла не более, чем 20-25\% длины полосы прокрутки даже в самом начале работы, когда статистика в интерполяционной таблице ещё не накоплена.

Сделать это эффективным образом можно в параллельном потоке выполнения, запускаемом после отображения грида. В этом потоке выполняется серия уточняющих запросов, по результатам которых пополняется интерполяционная таблица. Значение комбинации ключей для уточняющего запроса всякий раз выбирается как лежащее посередине самого большого зазора в значениях порядковых номеров записей интерполяционной таблицы. Процесс выполняется до тех пор, пока ширина максимального зазора не уменьшится до желаемого размера, либо до достижения ограничения на количество итераций.

\end{document}